\documentclass[aps,pre,twocolumn,showpacs,amsmath,amssymb,floatfix,superscriptaddress]{revtex4}
\usepackage{graphicx}
\usepackage{latexsym}
\usepackage{amsfonts}

\begin{document}

\title{The nonequilibrium discrete nonlinear Schr\"odinger equation}
\date{\today}

\author{Stefano Iubini}
\affiliation{Consiglio Nazionale delle Ricerche, Istituto dei Sistemi Complessi, 
via Madonna del Piano 10, I-50019 Sesto Fiorentino, Italy}
\affiliation{Dipartimento di Fisica e Astronomia, Universit\`a di Firenze and
INFN Sezione di Firenze, via Sansone 1, I-50019 Sesto Fiorentino, Italy}

\author{Stefano Lepri}
\email{stefano.lepri@isc.cnr.it}
\affiliation{Consiglio Nazionale delle Ricerche, Istituto dei Sistemi Complessi, 
via Madonna del Piano 10, I-50019 Sesto Fiorentino, Italy}

\author{Antonio Politi}
\affiliation{Institute for Complex Systems and Mathematical Biology \& SUPA
University of Aberdeen, Aberdeen AB24 3UE, United Kingdom}
\affiliation{Consiglio Nazionale delle Ricerche, Istituto dei Sistemi Complessi, 
via Madonna del Piano 10, I-50019 Sesto Fiorentino, Italy}

\begin{abstract}
We study nonequilibrium steady states of the one-dimensional discrete nonlinear
Schr\"odinger equation. This system can be regarded as a minimal model for
stationary transport of bosonic particles like photons in layered media or cold
atoms in deep optical traps. Due to the presence of two conserved quantities,
energy and norm (or number of particles), the model displays coupled transport 
in the sense of linear irreversible thermodynamics. Monte Carlo thermostats
are implemented to impose a given temperature and chemical potential at the
chain ends. As a result, we find that the Onsager coefficients are finite in
the thermodynamic limit, i.e. transport is normal. Depending on the position in
the parameter space, the ``Seebeck coefficient'' may be either positive or
negative. For large differences between the thermostat parameters, density and
temperature profiles may display an unusual nonmonotonic shape. This is due to
the strong dependence of the Onsager coefficients on the state variables.
\end{abstract}


\maketitle

\section{Introduction}

The Discrete Nonlinear Schr\"odinger (DNLS) equation 
\cite{Eilbeck1985,Kevrekidis}  has important applications  in many domains of
physics.  As it is well known, such equation arises in several  different
problems. A classical example is electronic transport in biomolecules
\cite{Scott2003}. In the context of optics or acoustics it  describes the
propagation of nonlinear waves in a  layered photonic or phononic system.
Indeed, in a suitable  limit, the dynamics of high-frequency Bloch waves is
described by a DNLS equation  for their envelope  (see Refs.
\cite{Kosevich02,Hennig99} for details). On the other hand, in the realm of the
physics of cold atomic gases, the equation  is an approximate semiclassical
description of bosons trapped in periodic  optical lattices (see e.g. Ref.
\cite{Franzosi2011} and references therein for a recent survey). Many other
physical problems have been recently  addressed having the DNLS equation as a
basic reference model, like the effect of nonlinearity on Anderson localization
\cite{Kopidakis2008, Pikovsky2008} and the violation of reciprocity in wave
scattering \cite{Lepri2011}  just to mention a few recent examples. 

While a vast literature has been devoted to localization problems, much less
is known about finite-temperature properties. The first analysis of the
equilibrium statistical mechanics of DNLS systems has been performed in
Ref.~\cite{Rasmussen2000}, while the relaxation of localized modes (discrete
breathers) in the presence of phonon baths has been discussed in
\cite{Rasmussen2000a,Rumpf2004}. Several results can be translated to
other types of nonlinear lattices, where a DNLS-like equation represents
an approximation of the lattice dynamics~\cite{Johansson2006}. 

An even less explored field is that of nonequilibrium properties of the
DNLS equation \cite{Eisner2006}. In particular, the case of an open system
that exchanges energy with external reservoirs has not been treated so far. 
The presence of two conserved quantitities naturally requires to argue about
coupled transport, in the sense of ordinary linear irreversible thermodynamics. 
Despite the very many studies of heat conduction in oscillator
chains~\cite{LLP03,DHARREV}, works on coupled transport are
scarce~\cite{Gillan85,Mejia2001,Larralde03}. Interest in this field has been
revived by recent works on thermoelectric phenomena~\cite{Casati2008,Casati2009}
in the hope of identifying dynamical mechanisms that could enhance
the efficiency of thermoelectric energy conversion~\cite{Horvat2009,Saito2010}.

In order to investigate transport properties, we need to introduce the
interaction of the system with external reservoirs that are capable to 
exchange energy and/or norm. For models like DNLS this is much less
straightforward than for standard oscillator chains, where 
e.g. Langevin thermostats are a typical choice \cite{LLP03,DHARREV}.
Here we propose and test a very simple Monte Carlo scheme which is easy to
implement and suitable for the model at hand.
Another important difference between the DNLS and standard oscillator chains
(like the Fermi-Pasta-Ulam or Klein-Gordon models) is that its Hamiltonian
is not the sum of kinetic and potential energies. Thus, it is necessary to
introduce suitable operative definitions of kinetic temperature $T$ and chemical
potential $\mu$, to measure such quantities in actual simulations. In the
following, we make use of a recent definition of the microcanonical temperature
\cite{Franzosi2011b} and extend it for the estimate of the chemical potential.

By imposing small $T$ and $\mu$ jumps across the chain, we can determine the
Onsager coefficients, which turn out to be finite in the thermodynamic limit,
i.e. both energy and mass conductions are normal processes. From the Onsager
coefficients we can thereby determine the ``Seebeck coefficient" $S$
\cite{note1} which we find to be either positive or
negative, depending on the thermodynamic parameters (i.e., mass and energy
density). For larger temperature or chemical-potential differences, although
one can still invoke the linear response theory, some surprising phenomena
emerge. One example is the ``anomalous heating" that can be observed when the
chain is attached to two thermostats operating at the same temperature: along
the chain, $T$ reaches values that are even three times larger than that imposed
on the boundaries. This phenomenon can be observed only in the case of coupled
transport, since it is due to the variable weight of the non-diagonal terms of
the Onsager matrix. It is apparent in the DNLS, because of the strong
variability of the Onsager coefficients.

The paper is organized as follows. In Sec.~\ref{sec:noneq} we introduce the
model and describe the heat baths. In Sec.~\ref{sec:obs}, we define the relevant
thermodynamic observables and the formalism (e.g., the Onsager coefficients)
necessary to characterize nonequilibrium steady states. Sec.~\ref{sec:steady}
is devoted to a discussion of the stead states, both in the case of small
and large $T$, $\mu$ differences. In Sec.~\ref{sec:zero} we provide a pictorial
representation of the general transport properties, by reconstructing the
zero-flux curves. Finally, the last section is devoted to the conclusions and to
a brief summary of the open problems.

\section{Setup}
\label{sec:noneq}

\begin{figure}[htbp]
\begin{center}
\includegraphics[width=0.4\textwidth,clip]{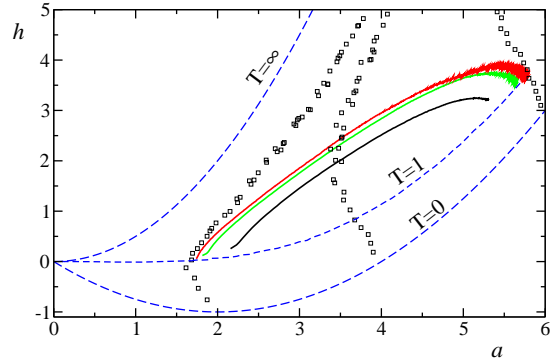}
\caption{(Color online) Parametric plots of the local norm and energies
$[a(y)$, $h(y)]$ for $T_L=T_R=1$, $\mu_L=0$, $\mu_R=2$ and increasing
chain lengths $N=200,800,3200$ (solid lines, bottom to top). The three
(blue) dashed lines
are the isothermal $T=0$, $T=1$ and $T=\infty$ respectively.
Lines at constant chemical potential (open symbols) $\mu=0$, $\mu=1$ and $\mu=2$
(left to right respectively)
are obtained by equilibrium simulations.
}
\label{Fig3}
\end{center}
\end{figure}

In one dimension, the DNLS Hamiltonian writes
\begin{equation}
H=\frac{1}{4}\sum_{i=1}^{N}\left(p_i^2+q_i^2\right)^2+
\sum_{i=1}^{N-1}\left(p_ip_{i+1}+q_iq_{i+1}\right)
\quad ,
\label{ham}
\end{equation}
where the sum runs over the $N$ sites of the chain. The sign of quartic term is
positive, as we refer to a repulsive-atom BEC, while the sign of the hopping
term is irrelevant, due to the symmetry associated with the canonical (gauge)
transformation $z_n \to z_ne^{i\pi n}$ (where
$z_n\equiv (p_n + \imath q_n)/\sqrt{2}$ denotes the amplitude of the wave
function). The equations of motion are
\begin{equation}
i\dot z_n = -z_{n+1}-z_{n-1}-2|z_n|^2 z_n
\label{dnlse} 
\end{equation}
with $n = 1, \cdots, N$, and fixed boundary conditions ($z_0=z_{N+1}=0$).
The model has two conserved quantities, the energy and the total norm (or total
number of particles) 
\begin{equation}
A=\sum_{i=1}^{N}(p_i^2+q_i^2) \quad .
\label{norm}
\end{equation} 
As a consequence, the equilibrium phase-diagram is two-dimensional, as it involves
the energy density $h=H/N$ and the particle density $a=A/N$. The first reconstruction
of the diagram was carried out in Ref.~\cite{Rasmussen2000} within the grand-canonical
ensemble with the help of transfer integral techniques. It is schematically described
in Fig.~\ref{Fig3}: the lower dashed line corresponds to the ground state ($T=0$) upon
varying the particle density; the upper dashed line corresponds to infinite
temperature. The nonequilibrium studies described in this paper correspond to
the region in between such two curves. 

We aim at characterizing the steady states of the chain when put in
contact (on the left and right boundaries) with two thermostats
at temperature $T_L$ and $T_R$ and chemical potentials $\mu_L$ and $\mu_R$,
respectively. The implementaton of the interactions with a heat bath is often
based on heuristics. In particle models, the simpler schemes consist in either
adding a Langevin noise, or in assuming random collisions with an equilibrium
gas \cite{LLP03,DHARREV}. For the DNLS this is less straightforward: adding white
noise and a linear dissipation drives the system to infinite temperature, i.e.
to a state in which relative phases are uncorrelated.
  
In the absence of a first-principle definition of heat bath, we consider two
phenomenological Monte-Carlo heat baths. The general scheme of this kind of
heat bath involves a stochastic dynamics which perturbs the canonical variables
$p_{1} \to p_{1} +\delta p_{1}$ and $q_{1} \to q_{1}+\delta q_{1}$ 
\cite{note2}
at random
times, chosen according to a uniform
distribution in the interval $[t_{min},t_{max}]$. The perturbations
$\delta p$ and $\delta q$ are independent random variables uniformly distributed
in the interval $[-R,R]$. Moves are accepted according to the standard Metropolis
algorithm, evaluating the cost function
$\exp{\{-T_L^{-1}(\Delta H -\mu_L\Delta A)\}}$ with $T_L$ and $\mu_L$
being the temperature and the chemical potential of the heat bath. 
This kind of thermostat exchanges both energy and particles. 
In some cases, however, we need to study the chain behavior in the absence of
one of the two fluxes (energy and norm). A simple way to study these setups
is to modify the perturbation rule of the thermostat, requiring the exact
conservation of the corresponding local density (energy density or norm
density). We have thus the following two schemes: 

\textit{Norm conserving thermostat-} The perturbation acts only on the phase
$\phi_{1}$ of the complex variable $z_{1}$. More precisely we impose
$\phi_{1}\to \phi_{1} + \delta \phi_{1} \ \ mod(2\pi)$, where
$\delta \phi$ is a random variable, uniformly distributed in the interval
$[0,2\pi]$. This dynamics conserves exactly the local amplitude $|z_{1}|^2$
and therefore the total norm $A$.

\textit{Energy conserving thermostat-}
In this case, it is necessary to go through two steps to conserve the
energy 
\begin{equation}
\label{e_dens}
e_1=|z_1|^4+2|z_1||z_2|\cos{(\phi_1-\phi_2)} \quad .
\end{equation}
First, the amplitude $|z_1|$ is randomly perturbed. As a result, both the local
amplitude and the local energy change. Then, by inverting, Eq.~(\ref{e_dens}),
a value of $\phi_1$ that restores the initial energy is seeked. If no such
solution exists, we go back to the first step and choose a new perturbation
for $|z_1|$.

There is a basic difference between the two types of thermostats. 
In the general scheme, a steady state is characterized by four parameters
$T_L$, $T_R$, $\mu_L$, $\mu_R$. On the other hand, for the norm-conserving
scheme we only assign $T_L$, $T_R$ and the norm density $a_{tot}$ of the whole
chain. As a consequence, the value of $\mu$ on the boundary is not fixed and
must be computed from the simulation. If the steady state is unique,
the former thermostating scheme must yield the same results, once the
chemical potentials are suitably fixed.
A numerical test of this equivalence has been performed, by reconstructing some
zero-flux profiles with both thermostats. The curves overlap reasonably
well, although some small systematic deviations are present.
This is because the norm flux is never exactly zero in the non-conservative
case (typically of order $\sim10^{-4}$ in a chain of 1000 sites). In addition,
there are slightly different thermal resistance effects in the two schemes. 
Besides those discrepancies, we conclude that the proposed schemes work equally
well for the generation of nonequilibrium steady states.

\section{Physical observables}
\label{sec:obs}

In order to characterize the thermodynamic properties of the DNLS, we extend
the approach of Ref.~\cite{Franzosi2011b} to derive an operative definition 
not only of the microcanonical temperature but also of the chemical potential.
The starting point are the usual definitions
$T^{-1}=\partial{\mathcal S}/\partial{H}$,
and $\mu/T=-\partial{\mathcal S}/\partial{A}$,
where $\mathcal S$ is the thermodynamic entropy. 
The partial derivatives must be computed taking into account the existence of 
two conserved quantities (hereafter called $C_1$ and $C_2$). Thus, 
\begin{equation}
\frac{\partial \mathcal S}{\partial C_1}=
\left\langle \frac{W\|\vec \xi\|}{\vec\nabla C_1\cdot\vec\xi}\ 
\vec\nabla\cdot\left(\frac{\vec\xi}{\|\vec\xi\|W} \right)\right\rangle
\label{tDnlse}
\end{equation}
where $\langle \: \rangle$ stands for the microcanonical average,
\begin{eqnarray}
\label{add}
\vec \xi&=&\frac{\vec{\nabla} C_1}{\|\vec{\nabla} C_1\|} -\frac{(\vec{\nabla}
C_1\cdot\vec{\nabla} C_2)\vec{\nabla}
C_2}{\|\vec{\nabla} C_1\|\|\vec{\nabla} C_2\|^2}
\\	
W^2&=&
\sum_{\genfrac{}{}{0pt}{}{j,k=1}{j<k}}
^{2N}\left[\frac{\partial C_1}{\partial x_j}\frac{\partial C_2}{\partial x_k}-
\frac{\partial C_1}{\partial x_k}\frac{\partial C_2}{\partial x_j}\right]^2
 \, , \nonumber
\end{eqnarray}
and $x_{2j}=q_j$, $x_{2j+1}=p_j$.
By setting $C_1=H$ and $C_2=A$, the above formula reduces to the expression for
$T$ derived in \cite{Franzosi2011b}. Moreover, by assuming $C_1=A$ and
$C_2=H$, Eq.~(\ref{tDnlse}) defines the chemical potential $\mu$.
Notice that both expressions are nonlocal. Nevertheless, we have verified that
it is sufficient to compute the expression (\ref{tDnlse}) over as few as 10 sites
to obtain, after some time averaging, reliable ``local'' estimates of both $T$ and
$\mu$ \cite{note3}.


The expressions for the local energy- and particle-fluxes are derived in the
usual way from the continuity equations for norm and energy densities,
respectively
\begin{eqnarray}
&& j_{a}(n) = 2\left(p_{n+1}q_n - p_nq_{n+1}\right) \\  
&& j_{h}(n) = -\left(\dot p_n p_{n-1} + \dot q_n q_{n-1}\right)
\label{flx}
\end{eqnarray}
The approach to the steady state is controlled by verifying that the 
(time) average fluxes are constant along the chain
($\overline{j_{a}(n)}=j_a$ and $\overline{j_{h}(n)}=j_h$).
Moreover it is also checked that $j_a$ and $j_h$ are 
respectively equal to the average energy and norm exchanged per 
unit time with the reservoirs.

As usual in nonequilibrium molecular dynamics simulations, some 
tuning of the bath parameters is required to minimize boundary 
resistance and decrease the statistical errors, as well as the 
finite-size effects \cite{LLP03}. For our Monte-Carlo thermostats,
we observed that it is necessary to tune the perturbation amplitude $R$.
Typically, there is an optimal value of $R$ for which one of the two currents
is maximal (keeping the other parameters fixed), but this value may depend on
$T$ and $\mu$. Since it would be unpractical to tune the thermostat parameters 
in each simulation, we decided to fix them 
in most of the cases. In particular we have chosen $R=0.5$, $t_{min}=10^{-2}$
and $t_{max}=10^{-1}$. Some adjustments have been made only when the fluxes
were very small.

In the thermodynamic limit (i.e. for sufficiently long chains), the local forces
acting on the system are very weak and one can thereby invoke the linear
response theory. This means that forces and fluxes are related by the relations
\cite{Saito2010}
\begin{eqnarray}
\label{lin}
j_a &=& -L_{aa} \frac{d (\beta\mu)}{dy} + L_{ah} \frac{ d \beta}{dy} \\
j_h &=& -L_{ha} \frac{d (\beta\mu)}{dy} + L_{hh} \frac{ d \beta}{dy} \nonumber
\end{eqnarray}
where we have introduced the continuous variable $y=i/N$, while $\beta$ denotes
the inverse temperature $1/T$; $\bf{L}$ is the symmetric, positive definite,
 $2\times 2$ Onsager matrix. 
Notice that the first term in the r.h.s. of the above equations is negative, since the 
thermodynamic forces are $-\beta\mu$ and $\mu$ and that 
${\rm det} {\bf L} = L_{aa}L_{hh}-L_{ha}^2>0$.

The particle ($\sigma$) and thermal ($\kappa$) conductivity can be expressed
expressed in terms of $\bf{L}$,
\begin{equation}
\sigma = \beta L_{aa};\quad \kappa = \beta^2 \frac{{\rm det}\bf{L}}{L_{aa}}\, .
\label{tcoeffs} 
\end{equation}
Analogously, the Seebeck coefficient $S$, which corresponds to (minus) the
ratio between the chemical-potential gradient and the temperature
gradient (in the absence of a mass flux), writes
\begin{equation}
S = \beta \left(\frac{L_{ha}}{L_{aa}} -\mu\right),
\label{seebeck} 
\end{equation}

We conclude this Section, by mentioning another important parameter, the figure of
merit
\[
ZT = \frac{\sigma S^2 T}{\kappa}=\frac{(L_{ha}-\mu L_{aa})^2}{detL};
\]
which determines the efficiency $\eta$ 
for the conversion of a heat current into a particle current
as \cite{Saito2010}
\[
\eta = \eta_C \frac{\sqrt{ZT+1}-1}{\sqrt{ZT+1}+1}. \qquad
\]
For large $ZT$, $\eta$ approaches the Carnot limit $\eta_C$.
Understanding the microscopic mechanisms leading to an increase
of $ZT$ is currently an active topic of research \cite{Casati2008}. 

\section{Steady states}
\label{sec:steady}

\subsection{Local analysis}

\begin{figure}[ht]
\begin{center}
\includegraphics[width=0.4\textwidth]{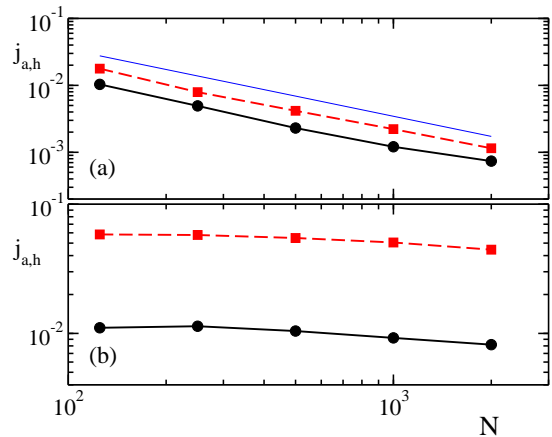}
\caption{Average energy current (squares) and norm current 
(dots) versus chain size $N$:
(a) High-temperature regime $T_L=2$, $T_R=4$, $\mu=0$ and 
(b) Low-temperature regime  $T_L=0.3$, $T_R=0.7$, $\mu=1.5$
The thin (blue) line is the $1/N$ behavior expected for normal
transport. Each value is obtained by computing the 
fluxes on a run of $5\times10^6$ time units. }
\label{jn}
\end{center}
\end{figure}

In a first series of simulations we have studied the nonequilibrium states
in the case of small differences between the two thermostats, verifying that
transport is normal, i.e.
the Onsager coefficients are finite in the thermodynamic limit. This is
less obvious than one could have imagined \cite{future}. In any case, for fixed
$\Delta T = T_R-T_L$ and $\Delta \mu = \mu_R-\mu_L$, the two fluxes
$j_a$ and $j_h$ are inversely proportional to the system size $N$. At high enough
temperatures, the asymptotic scaling sets in already in chains a few hundred sites long
(see Fig.~\ref{jn}a). Moreover, if $\Delta T$ $\Delta \mu$ and  are
small enough, the profiles of $T$ and $\mu$ along the chain are linear 
as expected. 

However, upon decreasing the temperature, the minimal chain length needed to
observe a normal transport, becomes very large. As shown in Fig.~\ref{jn}b, for
the same range of lattice sizes as in panel a, the currents are almost
independent on $N$, as one would expect in the case of ballistic transport.
This is because at small temperatures, one can always linearize the equations
of motion around the ground state (which depends on the norm density),
obtaining a harmonic description and thereby an integrable dynamics.

\begin{figure}[ht]
\begin{center}
\includegraphics[width=0.6\textwidth]{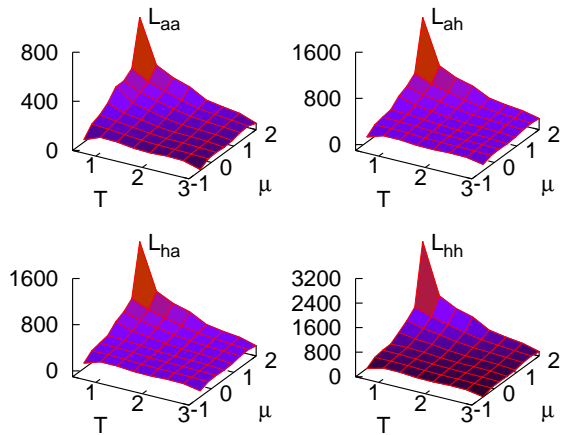}
\caption{Elements of the Onsager matrix $L$ in the $T,\mu$ plane. 
for a chain of length $N=500$; $\Delta T=0.1$, 
$\Delta \mu=0.05$. Each value is obtained by computing the 
fluxes on a run of $5\times10^6$ time units. }
\label{L}
\end{center}
\end{figure}

A plot of the four Onsager coefficients in the $(T,\mu)$ plane is 
reported in Fig.~\ref{L}. Within statistical errors, the off-diagonal 
terms are always positive in the considered range. All coefficients are
larger for small $T$ and large $\mu$. This is connected to the scaling
behaviour of the linear coefficients in the vicinity of the ground state
\cite{future}.

\begin{figure}[htbp]
\begin{center}
\includegraphics[width=0.4\textwidth]{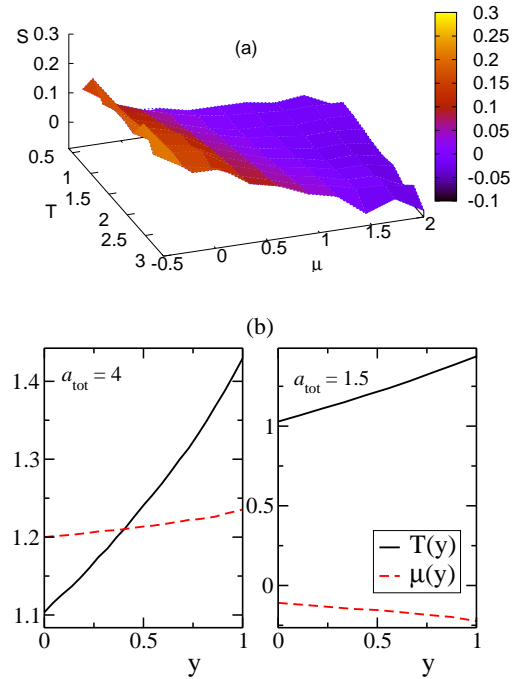}
\includegraphics[width=0.35\textwidth]{Fig4b.eps}
\caption{(Color online) (a) Seebeck coefficient $S$ obtained from the data in  
Fig.~\ref{L}; (b) Temperature and chemical protential profiles
for $T_L=1$, $T_R=1.5$; simulation with norm-conserving thermostats at two 
values of the norm density $a_{tot}$ corresponding to values of $S$ with
opposite signs.}
\label{fig:profs}
\end{center}
\end{figure}

The resulting coefficient $S$ is plotted in Fig.~\ref{fig:profs}a, where
one can see that there are two regions where the Seebeck coefficient is
positive, resp. negative, separated by a curve which, according
to Eq.~(\ref{seebeck}), is defined by $L_{ha}/L_{aa} =\mu$ (see below).
This means that the relative sign of the temperature and chemical-potential
gradients is opposite in the two regions (in the presence of a zero
norm-flux). This is indeed seen in Fig.~\ref{fig:profs}b where
the result of two different simulations are plotted in the two regions.

Finally, since the figure of merit $ZT$ roughly follows $S$, there is 
only a modest change in the considered parameter ranges.
Moreover, for fixed $T$, $ZT$ decreases upon increasing $\mu$.
This is qualitatively in agreement with the general expectation 
that an increasing strength of interaction (increasing $\mu$
means increasing the average norm and thus the nonlinearity)
is detrimental for the efficiency. 

\subsection{Global analysis}

If the temperature- or the chemical-potential difference is no longer small, the
temperature and chemical-potential profiles are expected to have a nonlinear
shape. This is because, as we have seen in the previous subsection, the Onsager
matrix varies with $a$ and $h$ (or, equivalently, with $T$ and $\mu$).

\begin{figure}[htbp]
\begin{center}
\includegraphics[width=0.4\textwidth]{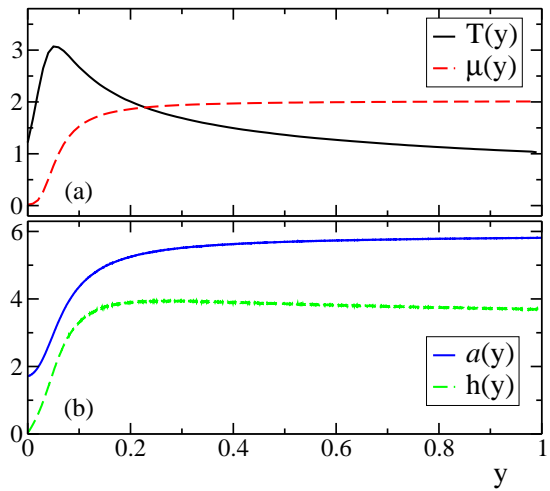}
\caption{(Color online) (a) Temperature and chemical potential profiles
as a function  of $y=i/N$ for a chain of $N=3200$ sites and $T_L=T_R=1$,
$\mu_L=0$, $\mu_R=2$. Each point is an
average of the appropriate microcanonical expression derived from
Eq.~(\ref{tDnlse}) over a subchain of about 10 sites around $i$.
(b) Norm and energy densities corresponding to the profiles in (a).}
\label{tmu}
\end{center}
\end{figure}

A particularly striking example is reported in Fig.~\ref{tmu}.
Both $T(y)$ and $\mu(y)$ profiles do approach the imposed
values at the chain edges (up to tiny jumps due to the boundary impedance).
However, $T(y)$ exhibits a remarkable non-monotonous profile: although the
chain is attached to two heat baths with the same temperature, it is
substantially hotter in the middle (up to a factor 3!).

Another way to represent the data is by plotting the local norm and energy 
densities in the phase plane $(a,h)$. By comparing the results for different
chain lengths, we see in Fig.~\ref{Fig3} that the paths are progressively
``pushed'' away from the $T=1$ isothermal and for $N=3200$ the asymptotic regime
is attained. 

In order to understand the onset of such anomalous shape, it is convenient to
rewrite Eq.~(\ref{lin}) by referring to $T$ and $\mu$. By introducing vector
notations, we can write,
\begin{equation}
{\bf J} = {\bf A}(\mu,T) \frac{d{\bf v}}{dy}
\end{equation}
where ${\bf J}= (j_a,j_h)$, ${\bf v}=(\mu,T)$, while the matrix ${\bf A}$ (which
is no longer symmetric) can be expressed in terms of the Onsager matrix and of
the fields $T$ and $\mu$ (for instance, $A_{11}= -L_{aa}/T$). By now inverting the
above equation one obtains
\begin{equation}\label{gen}
\frac{d{\bf v}}{dy} =  {\bf A}^{-1}(\mu,T) {\bf J}
\end{equation}
where ${\bf A}^{-1}$ denotes the inverse of $\bf A$. This system describes a set
of two linear differential equations which are non-autonomous 
(since the matrix coefficients in general vary with $\mu$ and $T$).

If one assumes to know the ``material'' properties (i.e. the matrix
${\bf A}^{-1}$) and wishes to determine fluxes and profiles, can proceed by
integrating the differential equations, starting from the initial condition
$T(0)=T_L$, $\mu(0)=\mu_L$. The, a priori unknown, parameters $j_a$ and $j_h$
can be determined by imposing that the final condition is $T(1)=T_R$ and
$\mu(1)=\mu_R$. Alternatively, if the fluxes are known, one can integrate
the equations up to any point $y_0$, and thereby generate the profiles that
would be obtained by attaching the right end of the chain to thermal baths
with temperature $T_R=T(y_0)$ and chemical potential $\mu_R=\mu(y_0)$.

\begin{figure}[htbp]
\begin{center}
\includegraphics[width=0.45\textwidth,clip]{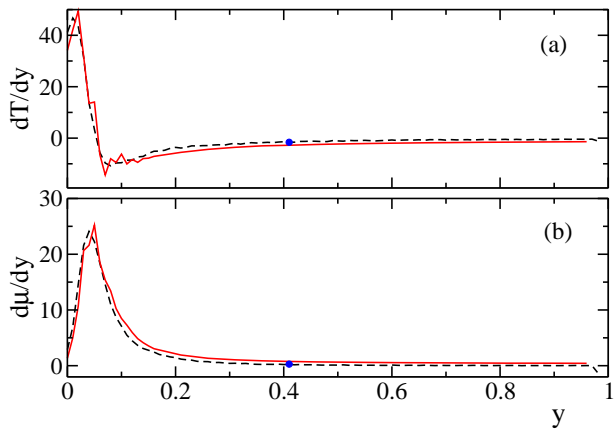}
\caption{(Color online) Spatial derivatives of $T$ and $\mu$
(panels (a) and (b), respectively)  computed from the profiles of Fig.~\ref{tmu}
 (black dashed lines) and their
reconstruction (red solid lines) by Eq.~(\ref{gen}) using the linear response
coefficients  (matrix $A$). The latter have been calculated on a chain of
$N=250$.  The quality of the recostruction improves by increasing the lattice
size has shown by tha blue filled dot which is obtained for $N=1000$. }
\label{der}
\end{center}
\end{figure}

In order to check the validity of the method, we have also adopted an
alternative point of view, by combining the knowledge of the fluxes with simulations
of short chains and small gradients to determine the elements of the matrix
$A$ in suitably selected points in the $(T,\mu)$ plane. In order to estimate the
four entries of $A$, it is necessary to perform two independent simulations
for, 
\begin{equation*}
\left\{
\begin{array}{ccc}
 T_{L,R} & =& T\pm \Delta T\\
\mu_{L,R} &=& \mu 
\end{array}
\right.
\ \ \ 
\left\{
\begin{array}{ccc}
 T_{L,R} & =& T\\
\mu_{L,R} &=& \mu \pm \Delta \mu 
\end{array}
\right.
\end{equation*}
With such information, we have been able to estimate $dv_i/dy$ along the chain 
(from Eq.~(\ref{gen})) and to compare the results with the direct
simulations. The results plotted in Fig.~\ref{der} demostrate that the two
approaches are in excellent agreement.


\section{Zero flux curves}
\label{sec:zero}

\begin{figure}[htbp]
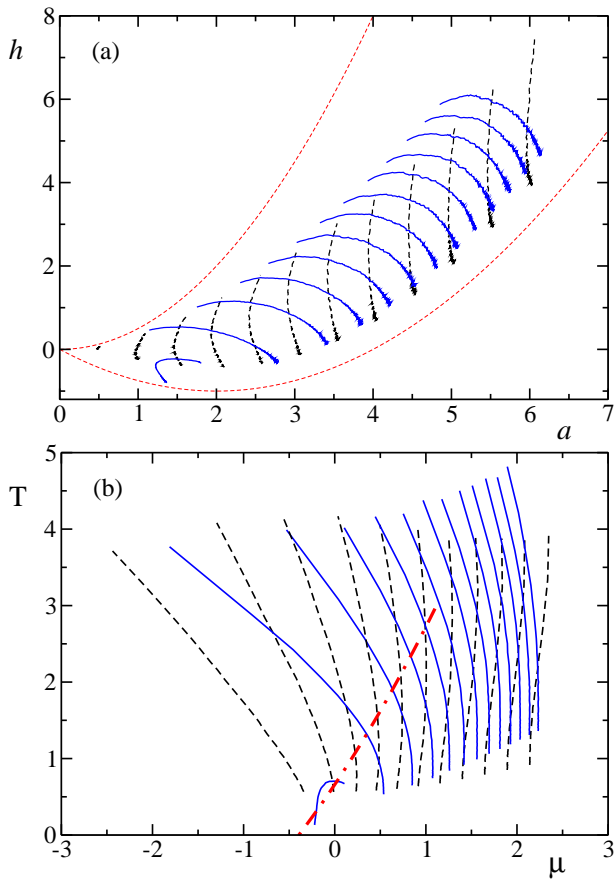

\begin{center}
\includegraphics[width=0.45\textwidth,clip]{Fig7a.eps}
\includegraphics[width=0.45\textwidth,clip]{Fig7b.eps}
\caption{(Color online) Zero-flux curves in the $(a,h)$ and 
$(\mu,T)$ planes (panels (a) and (b), respectively).
Black dashed and blue solid lines correspond to $j_a=0$ and 
$j_h=0$ respectively. Simulation are for a chain of length $N=500$. 
The thin dashed lines in the upper panel are the $T=0$ and $T=\infty$
isothermals. The thick dot-dashed line identify the locus where 
$S$ changes sign (see text).
}
\label{noflux}
\end{center}
\end{figure}

A compact pictorial representation of transport properties is obtained by
drawing the lines corresponding to vanishing fluxes $j_a$ and $j_h$. They can
be directly determined by means of the conservative thermostats presented
in Sec. II. Some lines are plotted in Fig.~\ref{noflux} both in the plane
$(a,h)$ and $(T,\mu)$. It is worth recall that in the absence of a mutual
coupling between the two transport processes (zero off-diagonal elements of the
Onsager matrix) such curves would be vertical and horizontal lines in the
latter representation. It is instead remarkable to notice that the solid
lines, which correspond to $j_h=0$ are almost veritical for large $\mu$:
this means that in spite of a large temperature difference, the energy flux
is very small. This is an indirect but strong evidence that the nondiagonal
terms are far from negligible.

The condition of a vanishing particle flux $j_a=0$ defines the Seebeck
coefficient which is $S= - d\mu/dT$. Accordingly, the points where the dashed
curves are vertical in Fig.~\ref{noflux}b identify the locus where $S$ changes
sign. The $j_h=0$ curves have no direct interpretation in terms of 
standard transport coefficients. Finally, if one connects a DNLS chain
with any two points in the $(\mu,T)$ plane, its profile would correspond
to the only path that is characterized by a constant ratio of $j_a/j_h$.

It is instructive to compare these results with the scenario expected in the
``harmonic'' limit, where the nonlinear terms in the DNLS are negligible. 
Here, the dynamics is characterized by an ensemble of 
freely propagating waves and transport is thus ballistic. A direct
reconstruction of the zero-flux lines by direct simulations is not very useful,
as, in analogy with the known behaviour for the harmonic chain \cite{RLL67}, the
profiles of $T$ and $\mu$ are flat (except for a few sites close to the
boundaries). Thus, the curves degenerate to single points and no comparison is
possible. We thus resort to a different method of computing transport
coefficients for ballistic systems, which is completely analogous to the
well-know Landauer theory of electronic transport~\cite{Sheng2006}.
Consider an $N$-site chain
in between two leads at different temperatures and chemical potentials 
$(T_L,\mu_L)$, $(T_R,\mu_R)$. Since transport is ballistic, energy and norm are
carried  by $N$ independent phonon modes, whose dispersion law is 
$\omega(q) = 2 \cos q$, $q$ being the wavenumber ($|q|\le \pi$). 
Accordingly, the fluxes are $N$-independent and the ensuing transport
coefficients are proportional to $N$. In this context, the norm and energy
fluxes are given (up to some numerical constant) by the formulae 
\begin{eqnarray*}
J_a   &=& \int_{-2}^{+2}\, d\omega \, t(\omega) [f_L(\omega) - f_R(\omega)]\\
J_{h} &=& \int_{-2}^{+2}\, d\omega \, \omega \, t(\omega) \, 
[f_L(\omega) - f_R(\omega)]\, , 
\end{eqnarray*}
where $t(\omega)$ denotes the transmission coefficient, while
$f_{L,R}$ are the equilibrium distribution functions of the 
reservoirs. If we assume that they are composed of two infinite 
linear chains (both with the same dispersion), the equipartition 
principle implies that the distributions
are of the Rayleigh-Jeans form \cite{Rumpf2004}, 
$f_{L,R}(\omega)= f(T_{L,R},\mu_{L,R},\omega)$ where
\[
f(T,\mu,\omega)=\frac{T}{\omega-2\mu}\quad,
\]
(the factor 2 stems from the definition of $z_n$ and Eq.~(\ref{norm}))  
The physical meaning of the formulae is pretty intuitive: they can be derived
from suitable generalized Langevin equations \cite{future} following
similar steps as for coupled oscillators,
see e.g. Ref.~\cite{DR06}. The relevant information is contained in the 
transmission coefficient, that depends on how the chain is coupled
to the external leads. For the Monte-Carlo bath we have used throughout this
paper, the precise form of $t$ is not known. We thus postulate the simplest
possible form, namely that, for large $N$, $t(\omega)=t$ for $|\omega|<2$ and
zero otherwise. For our purposes, we set $t=1$ in the following, otherwise all
the coefficients must be multiplied by $t$. If we introduce the function
\[
\Phi(T,\mu) \equiv \int_{-2}^{+2}\, d\omega \, f(T,\mu,\omega)
=T\ln\left(\frac{\mu-1}{\mu+1}\right)
\]
which for $\mu<-1$ and $T>0$ is always positive, we can write,
\begin{eqnarray*}
J_a   &=& \Phi(T_L,\mu_L)-\Phi(T_R,\mu_R)\\
J_{h} &=& 4(T_L-T_R)+
2\mu_L \Phi(T_L,\mu_L) - 2\mu_R \Phi(T_R,\mu_R) \, .
\end{eqnarray*} 
By expanding to first order in $\Delta T=(T_L-T_R)$ and 
$\Delta\mu=(\mu_L-\mu_R)$ 
\begin{eqnarray}
J_a   &=& M_{00}\Delta T + M_{01}\Delta \mu \nonumber \\ 
J_{h} &=& M_{10}\Delta T + M_{11}\Delta \mu \,  
\label{jland}
\end{eqnarray} 
where 
\begin{equation*}
 M = \left(
\begin{array}{cc}
\Phi(T,\mu) & \frac{2T}{\mu^2-1} \\
4+2\mu\Phi(T,\mu) & \frac{4T\mu}{\mu^2-1}+2T\Phi(T,\mu) \, 
\end{array}
\right).
\end{equation*}
With the help of the explicit formulae (\ref{jland}) we can reconstruct 
the zero-flux curves as follows. Starting from an initial point 
$(T_{in},\mu_{in})$ we compute $\Delta T$ and $\Delta\mu$ inverting 
Eqs.~(\ref{jland}) setting $J_a=0, J_h=1$ and $J_a=1, J_h=0$, respectively
(the value 1 is arbitrary). We then let 
$(T_{in},\mu_{in}) \to (T_{in}+\Delta T,\mu_{in}+\Delta\mu)$ and iterate
the procedure until the whole lines are reconstructed.

\begin{figure}[ht]
\begin{center}
\includegraphics[width=0.4\textwidth,clip]{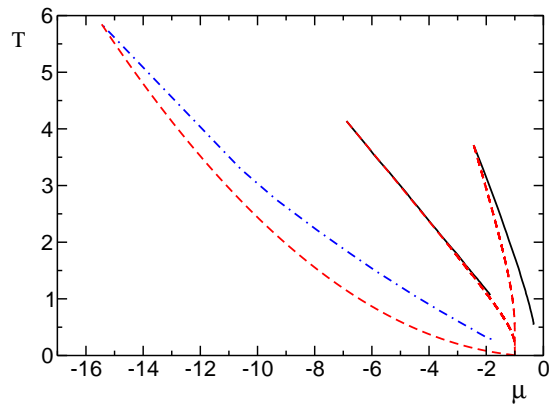}
\caption{(Color online) Comparison of the zero-flux lines 
obtained from simulation of the DNLS equations (black
 solid and blue dot-dashed lines  
correspond to $j_a=0$ and $j_h=0$ respectively). 
The dashed (red) lines are the zero-flux lines computed by the 
Landauer formulae as described in the text.
}
\label{fig:land}
\end{center}
\end{figure}

The results are depicted in Fig.~\ref{fig:land}. The curves for 
the linear case are defined only in the region $\mu<-1$. 
The results of the simulations of the DNLS (solid lines) 
nicely approach the curves of the linear case (dashed lines) upon decreasing 
$\mu$. The agreement is satisfactory, expecially in view of the 
many ad hoc assumptions made in deriving Eqs.~(\ref{jland}).

\section{Conclusions}

We have presented the first study of stationary transport properties 
of the DNLS equation. Due to the nonstandard form of its Hamiltonian, 
several new issues have been brought to the fore that had not been
addressed before in the literature dealing with energy transport 
in oscillator chains. In particular, we have extended the microscopic
definition of the temperature to the chemical potential and defined suitable
Monte Carlo thermostating schemes to characterize the nonequilibrium steady
states of the DNLS in various regimes. The simulations confirm the expectations
that transport is normal (i.e. the Onsager coefficients are finite in the 
thermodynamic limit), although some almost ballistic transport is found at very
low temperature, where the DNLS approaches an almost integrable limit. 

Due to the very existence of two naturally coupled transport processes, 
the nonequilibrium steady state can display nonmonotonous energy and
density profiles. To our knowledge, this unusual feature has not been observed
so far in any other oscillator or particle model.
As seen from Eq.~(\ref{gen}), it is clear that the temperature profile cannot in
general be linear in $y$, since the elements of ${\bf A}^{-1}$ depend on 
$\mu$ and $T$. In principle, the profiles may have have nontrivial shapes 
depending on the qualitative behaviour of the solutions of Eq.~(\ref{gen}).
In the DNLS, the phenomenon
is particularly pronounced (the temperature inside the chain reaches values
that are almost three times larger than those imposed by the thermal baths)
because of the strong variability of the Onsager coefficients. It would be
interesting to find the physical motivation for this effect to predict
and possibly control the conditions for its appearance.

Another novel feature is the fact that the Seebeck coefficient changes sign
upon changing the state parameters e.g. by increasing the interaction strength.
The observable consequence of this is that the 
temperature and chemical potential gradients change their relative
signs. As the particle density $a$ increases with $\mu$ this also
implies that $a$ may be larger in the colder regions.

Furthermore, a remakable feature of the DNLS thermodynamics is the possibility
of negative temperatures states in suitable parameter
regions~\cite{Rasmussen2000}. These regions, that are characterized by the
presence of long-lived localized excitations (discrete breathers), have not be
considered in the present paper, but are definitely worth being explored. 
It may be indeed speculated that they would lead to genuine nonlinear
transport features and even to the birth of new dynamical regimes 
possibly displaying transitions between conducting and insulating 
states.

Besides its intrinsic theoretical interest as a testbed for the characterization
of coupled irreversible processes, the DNLS equation opens the way also to
experimental investigations. In fact, despite its mathematical
simplicity, the DNLS model can be of guidance in the design and interpretation
of experiments on coupled-transport in cold atomic gases in deep optical
lattices as well as in optical multilayered and nonlinear structures. 

\acknowledgments

We thank R. Livi and C. Mej\'ia-Monasterio for fruitful 
discussions. This work is part of the Miur PRIN 2008
project \textit{Efficienza delle macchine termoelettriche: 
un approccio microscopico}.


\begin{thebibliography}{28}
\bibitem{note1}With a slight abuse of
terminology, we use the language of thermoelectric phenomena, even though
the underlying physical process is thermodiffusion, since particles have no
electric charge in the DNLS context.
\bibitem{note2} For
the sake of simplicity we refer to the left boundary, but the same rule
refer to the $N$th site as well.
\bibitem{note3} In general, both $T$ and $\mu$ contain additional
terms $\tilde T$ and $\tilde \mu$ that can, however, be neglected.
In the DNLS case, it can be proven that $\tilde T\equiv 0$, while 
simulations indicate that $\tilde \mu \neq 0$ but 
also that $\langle\tilde \mu \rangle = 0$. Indeed, the value of $\mu$ 
computed ignoring $\tilde \mu$ coincides 
with the expected value when $\mu_L=\mu_R$, $T_L=T_R$. In any case,
even in a nonequilibrium setup,
$\mu$ must be computed on sufficiently long subchains. 
This choice is particularly delicate for small nonlinearities and low temperatures. 
For instance, for $T=0.5,\mu=-0.5$ subchains of at least $10 \div 20$ sites 
are needed.
\end{thebibliography}

\end{document}